\documentclass[aps, reprint, longbibliography]{revtex4-2}
\usepackage{graphicx, natbib, amsmath}
\usepackage{xcolor} 
\usepackage{subfigure, amssymb}
\usepackage{float} 
\usepackage{color}

\DeclareGraphicsExtensions{.pdf,.eps,.png,.jpeg,.mps}
\graphicspath{ {Figures/} }

\begin{document}

\title{Metal Assisted Chemical Etching patterns at a Ge/Cr/Au interface \\ 
modulated by the Euler instability }
\author{Yilin Wong}
\author{Giovanni Zocchi}
\email{zocchi@physics.ucla.edu}
\affiliation{Department of Physics and Astronomy, University of California - Los Angeles}

\begin{abstract}
\noindent We present a solid state system which spontaneously generates remarkable engraving patterns on 
the surface of Ge. The layered construction, with a metal film on the Ge surface, results in coupling of the metal catalyzed 
etching reaction with the long range stress field at the Ge - metal interface. The etching patterns generated have 
similarities with Turing patterns, hydrodynamic patterns, crack propagation, and biological form. We describe spirals, radial patterns, 
and more disordered structures. Euler buckling of the metal layer generates a characteristic wavelength for some patterns.  
\end{abstract}


\maketitle


\section{Introduction} 

\noindent Understanding patterns in nature is a basic quest of science \cite{Cross_Book}. While biological form 
\cite{Darcy_Thompson_Book} inspired the pioneering work of Alan Turing \cite{Turing1952} on 
reaction - diffusion systems, most nonlinear dynamics studies that followed focussed on simpler 
experimental systems, such as thermal convection or crystal growth \cite{Cross1993, Langer1980, Gallaire2017}. 
Here we present a solid state system which, starting from homogeneous initial conditions, generates remarkable patterns 
at the $10 - 100 \, \mu m$ scale. The patterns arise from a complex interplay of solid state mechanics, chemical reactions, 
and diffusion. Although the system is very simple, its behavior is complex. It intersects several topics in nonlinear 
physics and engineering. The basis for the patterns is Metal Assisted Chemical Etching (MACE \cite{Huang2011}), 
combined with thin film mechanical instability \cite{Bowden1998}. \\ 
Initially shown to produce porous Si for optoelectronics applications \cite{Li2000}, 
MACE is a relatively recent technique pursued for high throughput fabrication of 3D structures 
on semiconductor surfaces \cite{Hildreth2009, Kawase2013, Huang2011}. 
It consists of depositing a noble metal, 
in the form of nano particles or a thin film, onto the surface of Si or Ge. In the presence of oxidizing agents such as $HF$ 
or $H_2 O_2$ the metal catalyzes etching of the semiconductor, creating 3D patterns such as deep wells \cite{Hildreth2009}. 
By shaping the metal catalyst, it is possible to design specific etching patterns, such as spirals \cite{Hildreth2012}. 
These applications take advantage of the metal catalyst to localize and direct the etching reaction. 
3D structures can be obtained because the metal catalyst, e.g. in the form of nano particles, ``travels'' with the etching 
front \cite{Rykaczewski2011}. \\ 
The buckling of thin metal films under compressive stress has been studied extensively in the past few decades 
due to its relevance to diverse fields such as flexible electronics \cite{McAlpine2011}, design of structural 
sandwich panels \cite{Sandwich_Book} and the understanding of mechanical instabilities in morphogenesis \cite{Nelson2016}. 
Understanding the instabilities of thin films under tensile stress is equally important, with recent work 
highlighting different regimes of the wrinkling instability \cite{Cerda2011} and addressing patterns formed by  
crack propagation in the film \cite{Marthelot2014}. \\
Our system is prepared from a p-doped, $\sim 200 \, \mu m$ thick Ge wafer (see Mat. \& Met. in  Supp. Mat. for details). 
A $\sim 10 \, nm$ thick Cr layer is evaporated 
on the (100) surface of the chip, followed by a $4 \, nm$ Au evaporation. To start the process, a drop of mild etching solution  
is transferred on the metallized surface and allowed to dry overnight. After washing off precipitated salt, the chip is re-incubated 
with the same etching solution in a wet chamber to prevent evaporation. From these homogeneous initial conditions, 
in the course of 24 - 48 hrs, remarkable patterns emerge on the Ge surface, revealed by the metal layer lifting off. 
The long range elastic stress field in the metal film couples with catalysis at the Ge - metal interface and diffusion 
of the etching solution to provide a mechanism for instability which can result in ordered large scale patterns. 
With coupled local (diffusion and reaction controlled) and global (elastic field controlled) growth mechanisms, 
the patterns obtained have a Turing - like as well as a hydrodynamic character. 
The thickness of the metal layer, the initial state of mechanical stress of the sample, 
and the composition of the etching solution all play a role in determining the type of pattern that develops.\\

\section {Experimental Observations} 
\begin{figure}[H]
	\centering
	\subfigure[Archimedean][\label{fig:Archimedean}]{\includegraphics[width=\linewidth]{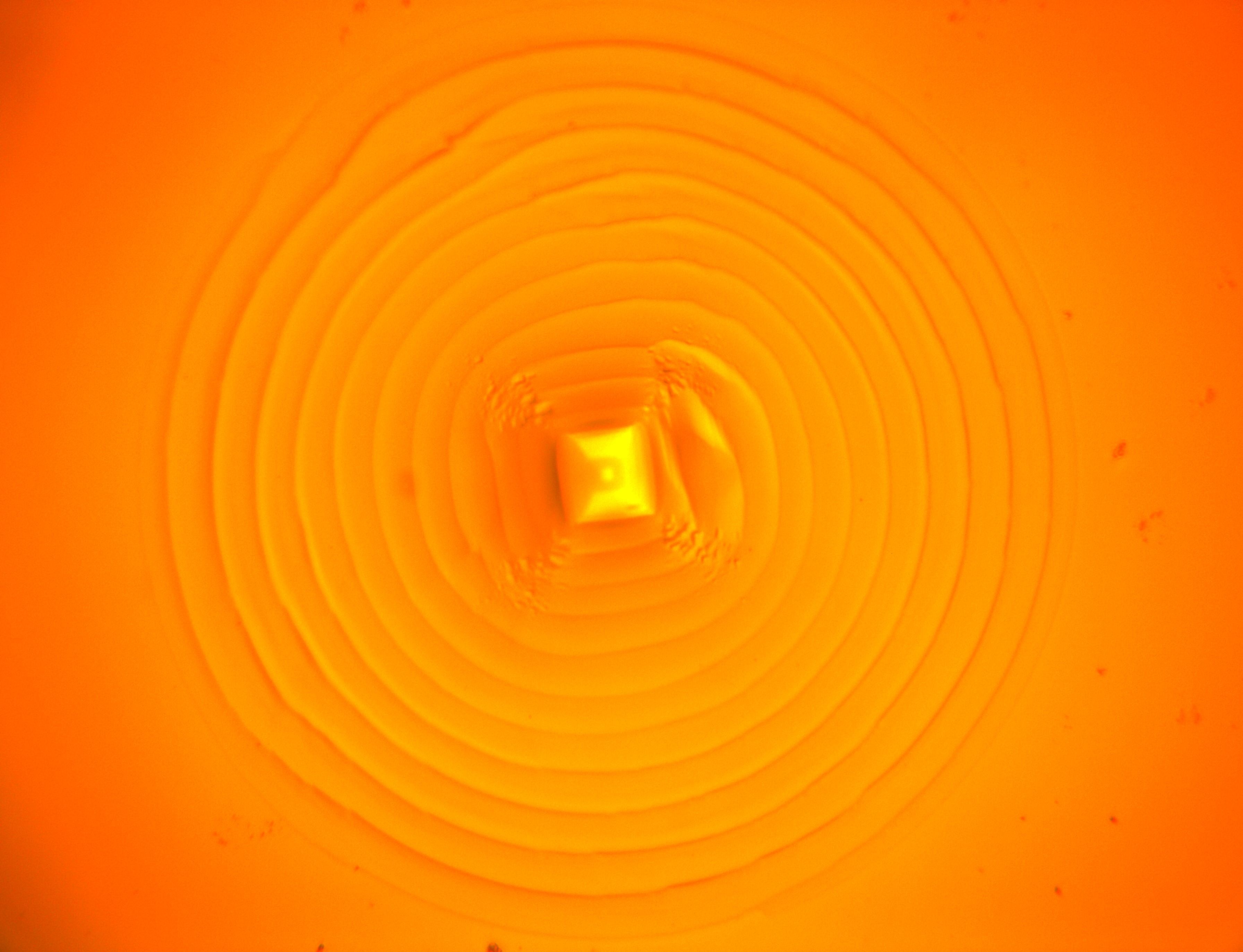}}
	\subfigure[Logarithmic][\label{fig:Logarithmic}]{\includegraphics[width=\linewidth]{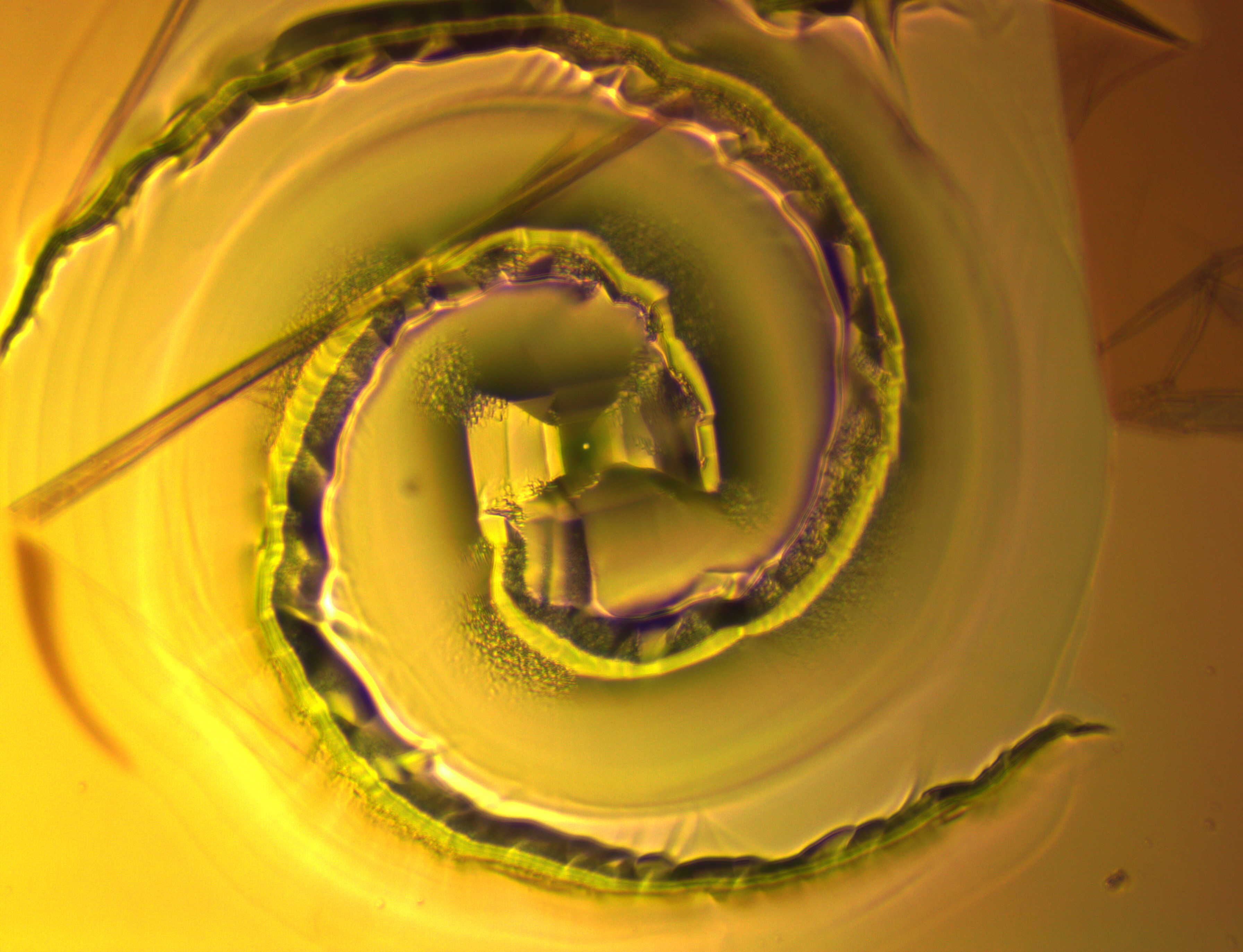}}
	\caption{(a) Archimedean spiral formed on Ge surface with $10 \, nm$ Cr and $4 \, nm$ Au depositions  
	($1 M \, KH_2PO_4$ buffer at $pH \, 4$ with $10 \, \mu M$ DNA). The square etch pit at the center is $9 \, \mu m$ in lateral size 
	and $ \sim 10 \, \mu m$ deep. Two etch ridges (spiral ``arms'') sprout from the NE corner of the etch pit, 
	forming a system of two intertwined ascending spiral terraces. The width of a terrace is $10 \, \mu m$ and the outer diameter 
	of the structure is $200 \, \mu m$. \\ 
		(b) Logarithmic spiral formed with $6 \, nm$ Cr and $4 \, nm$ Au on Ge, same etch solution as in (a). 
		The pattern is $600 \, \mu m$ in diameter. The diagonal line across the NW corner of the picture 
		is a rolled up piece of metal film detached from the Ge base. }
	\label{fig:primary}
\end{figure}

\begin{figure}[H]
	\centering
	\includegraphics[width=3.0in]{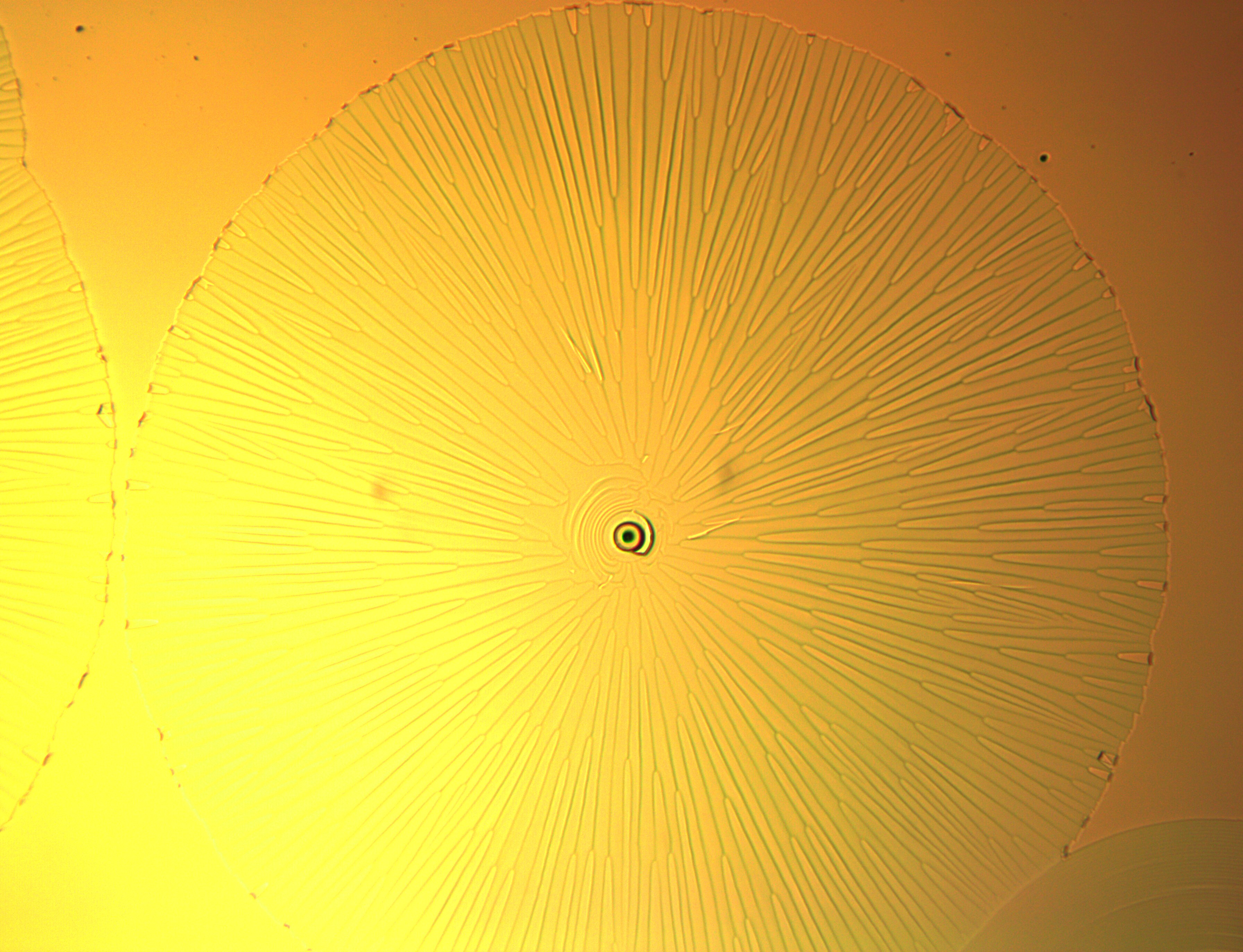}
	\caption{Radial pattern of grooves (dark lines) formed on Ge with $20 \, nm$ Cr and $4 \, nm$ Au ($10 \, mM$ TCEP,  
	$KH_2PO_4$ buffer, no DNA). The metal layer has lifted off the surface of the disk, which is $2 \, mm$ in diameter. 
	Tip splitting maintains a constant wavelength of the pattern (distance between grooves) of $\ell \approx 20 \, \mu m$ .}
	\label{fig:Radial}
\end{figure}

\begin{figure}[H]
	\centering
	\includegraphics[width=3.0in]{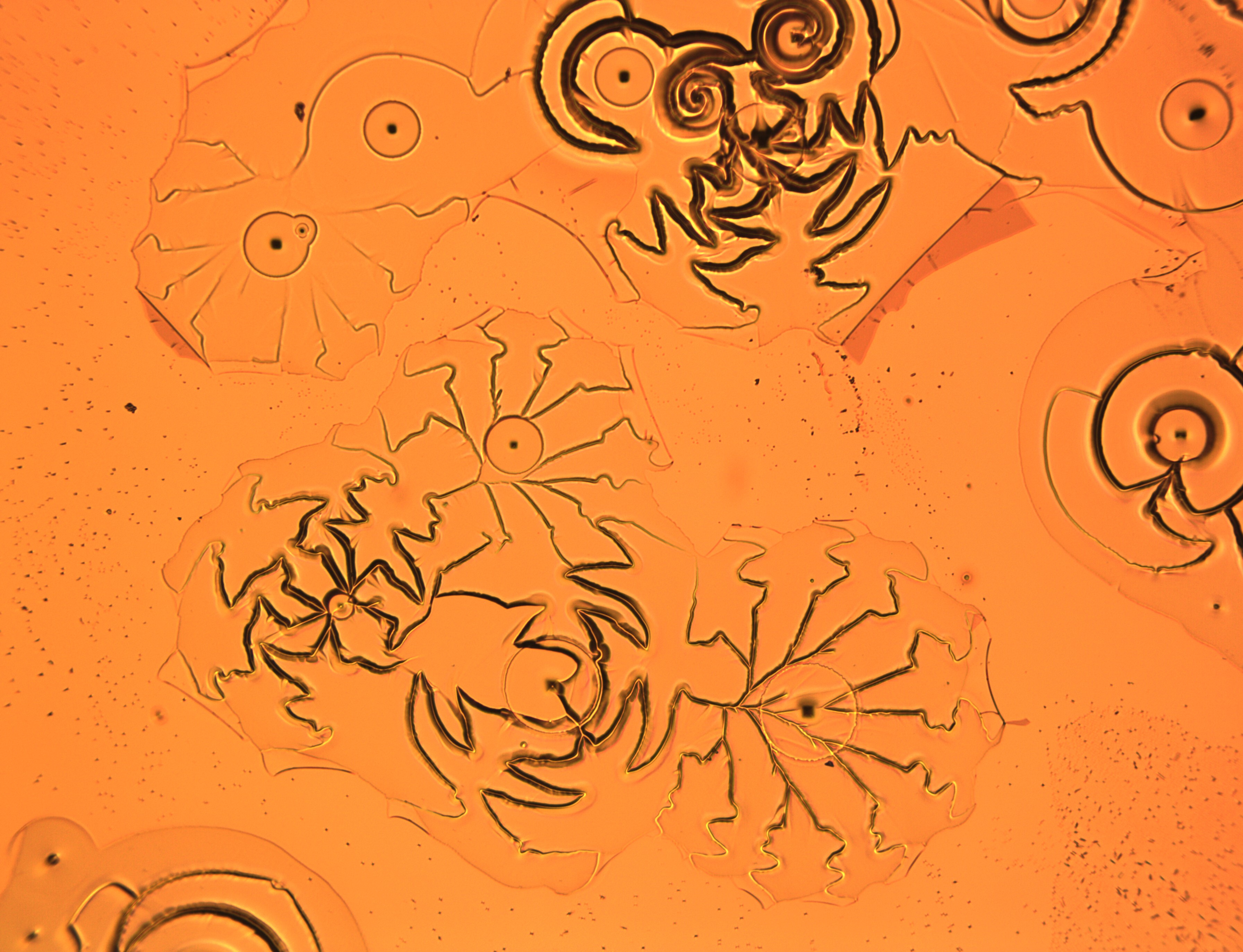}
	\caption{``Scary patterns'' formed on Ge with $3 \, nm$ Cr and $4 \, nm$ Au.}
	\label{fig:scary_pattern}
\end{figure}

\raggedbottom
\noindent The system is prepared as a laterally homogeneous solid state construct (i.e. with no pre-imposed pattern) 
consisting of the Ge substrate, an evaporated layer of Cr, and a further layer of Au (Mat. \& Met.). Then a mild etching 
solution, basically slightly acidic water ($H_2KPO_4$ buffer at $pH = 4$), is added. Over a period of $\sim 24 - 48$ hrs, 
a variety of stunning geometrical surface patterns at the $\sim 100 \, \mu m$ scale appear. These patterns are etched 
onto the Ge surface, at the interface with the metal film; the latter eventually lifts off the Ge surface 
during the etching process. For given conditions, 
one pattern, say spirals, is dominant or even exclusively present, appearing at seemingly random locations, so that on a chip 
$\sim 1 \, cm^2$ in size there may be hundreds of nearly identical spirals. There appear to be two ``fundamental'' 
geometric patterns: the spiral (Fig. \ref{fig:primary}) and the radial pattern (Fig. \ref{fig:Radial}). The spirals are further 
subdivided into Archimedean (Fig. \ref{fig:Archimedean}) and Logarithmic (Fig. \ref{fig:Logarithmic}). 
We can also generate other, seemingly disordered patterns (Fig \ref{fig:scary_pattern}): we believe these 
correspond to conditions ``in between'' those giving rise to either spirals or radial patterns, possibly also in the presence of  
gradients in the system (e.g. temperature gradients). The basic patterns always originate from a central ``etch pit'' 
and grow outwards. Fig. \ref{fig:primary} shows pictures obtained by reflection optical microscopy of the dry sample; 
the square structure at the center of the spiral is the etch pit. It has an inverted pyramidal structure with flat bottom and 
terraces, and aspect ratio $\sim 1$. The lateral size is $2 - 10 \, \mu m$ in our samples, with similar depth. The sides 
of the square are always alligned with the lattice axes of the (cubic) Ge crystal, as we see from the fact that all etch pits are 
oriented the same way on the [100] surface of the Ge wafer. Fig. S5 (Supp. Mat.) shows an SEM picture of the etch pit. 
From the etch pit, $2 - 6$ ``arms'' may sprout out, typically at the corners of the square. In Fig. \ref{fig:Archimedean}, 
the terraced square structure of the pit ``sprouts'' two arms from the NE corner; these become  
the ridges of the two-armed tight spiral in the figure. 
Note that, contrary to appearances, the pattern in the figure forms a valley, not a hill, on the Ge surface. As one walks 
(clockwise) along one or the other of the two spiral terraces, moving outwards from the center, one is ``climbing'' 
towards the level of the unetched surface of the chip. Indeed the maximum diameter of the pattern ($\sim 200 \, \mu m$ in 
Fig. \ref{fig:Archimedean}) corresponds to the spiral terraces reaching the level of the unperturbed 
Ge surface. The step separating one terrace from the adjacent one is of order $1 \, \mu m$. In Fig. \ref{fig:Archimedean} we 
count 10 terraces in the radial direction; the overall depth change, measured approximately by focussing the microscope 
alternately near the center and at the rim, is approximately $10 \, \mu m$. 
In the picture, the metal film has lifted off the Ge surface. In general, as the etching reaction proceeds, the metal film 
corrugates and eventually lifts off the Ge surface. 
In the spiral of Fig. \ref{fig:Archimedean}, the distance between the two arms (i.e. the width of the terraces) 
is constant as one moves radially outwards. Geometrically this corresponds to a curve, in polar coordinates $ (r , \theta )$, 
given parametrically by $\{ r = u t \, , \, \theta = \omega t \, , \, 0 \le t \le t_{max} \}$ (Archimedean spiral). The distance 
between the same arm at two successive turns is $\Delta r = 2 \pi u / \omega$ , independent of $r$. In Fig. \ref{fig:Archimedean}, 
this distance is twice the width of one terrace. It is indeed constant in the experiments, and also approximately the same 
across spirals on the same sample.  \\ 

The pattern of Fig. \ref{fig:Logarithmic} is obtained with only a slight change in parameters (thinner Cr layer). It has the same 
topology as the pattern in Fig. \ref{fig:Archimedean}: the two ``arms'' of the spiral are ridges delimiting two terraces 
which are sloping upwards as one turns clockwise, moving outwards from the central pit. 
However, the arms now follow an approximately 
logarithmic spiral shape (Supp. Mat. Fig. S3), the width of the terraces increasing moving outwards. The pattern is more jagged 
compared to the smoother Archimedean spiral, and the ridges are deeper. In polar coordinates, 
the parametric form of the logarithmic spiral is $\{ d r / d t = r /\tau \, , \, \theta = \omega t \, , \, 0 \le t \le t_{max} \}$ , 
i.e. $r = r_0 \,  e^{\theta / (\omega \tau) }$ . For the pattern of Fig. \ref{fig:Logarithmic} , $\omega \tau \approx 3.8$ 
(Supp. Mat. Fig. S3). \\
Since etching involves primarily a redox reaction (see Discussion), we guessed that addition of an oxidizing or reducing agent 
could alter the dynamics and the patterns obtained. Indeed, Fig. \ref{fig:Radial} shows a quite different, radial pattern 
obtained with the addition of TCEP (a reducing agent). The figure is an optical microscope picture of 
the dry sample; the metal layer has lifted off the area of the circular disk, exposing the Ge surface. 
The pattern of grooves (darker lines) 
has on average a constant wavelength $\ell$ (distance between grooves), maintained by ``tip splitting'' of the grooves as the pattern 
grows radially. For the pattern of Fig. \ref{fig:Radial} , $\ell \approx 22 \, \mu m$ , while the diameter of the disk is 
$\approx 2 \, mm$ . Like the spirals of Fig. \ref{fig:primary} , the radial pattern also starts from an etch pit at the center. 
Immediately surrounding the pit there is a circular pattern, which then transitions to the radial pattern (Fig. \ref{fig:Radial}). 

\noindent Through time-lapse video microscopy of samples under water we have obtained preliminary measurements 
of the growth dynamics of the basic patterns. Both the Archimedean spiral and the radial pattern have constant radial growth rate $u$ , 
of approximately $u = dr / dt \approx 0.75 \, \mu m / min$ for the sample measured (Supp. Mat. Figs. S6, S7). 
The Archimedean spiral ``turns'' at constant angular velocity 
$\omega$ of approximately $6 \, ^{\circ} / min$ (Fig. S9) ; this gives a wavelength of $2 \pi u / \omega \approx 45 \, \mu m$. 
Finally, besides the regular spiral and radial patterns, more disordered - and quite arresting - patterns can be obtained, 
under conditions which we do not entirely control (Fig. \ref{fig:scary_pattern}). These appear to be combinations of radial and 
spiral elements, possibly indicating that conditions are close to a bifurcation point separating two different regular patterns. 

\begin{figure}[H]
	\centering
	\includegraphics[width=3.5in]{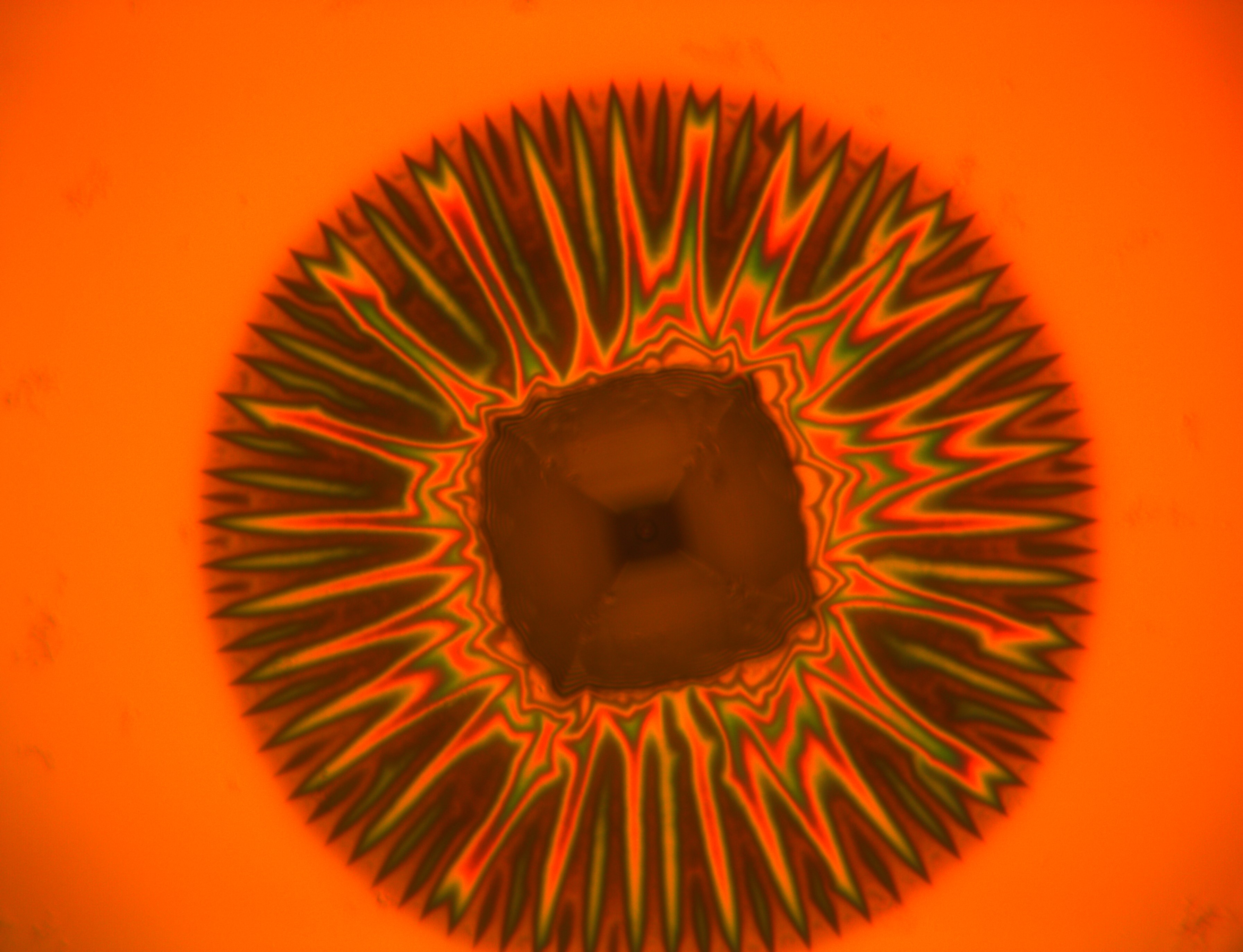}
	\caption{Interference microscopy picture of a radial structure in water, before the metal film lifts off. 
	Light reflection from the Ge surface and the metal layer forms the colorful interference fringes showing the corrugations  
	of the metal layer. The process gives rise to a pattern of grooves on the Ge as in Fig. \ref{fig:Radial}. The diameter of the disk 
	is $22 \, \mu m$. }
	\label{fig:interference}
\end{figure}

\section  {Discussion} 

\noindent We observe a stunning variety of patterns produced by metal catalyzed etching of the Ge (100) surface. The patterns 
always start from an initial etch pit and propagate outwards. The etch pit and surrounding tight spiral (Fig. 1a) are 
well known patterns, thought to be associated with preferential etching at a screw dislocation \cite{Rhodes1957}. 
The association of etch pits with screw dislocations is well established, indeed the former have been used to measure 
density and dynamics of the latter \cite{Amelinckx_Book}. 
The mechanism of formation of the surrounding spiral is less established; several authors note that 
if etching of the spiral was a time reversed crystal growth process along the screw dislocation, the dislocation 
in question would have an enormous ($\mu m$ size) Burgers vector, which seems implausible \cite{Rhodes1957}.  
For our system, the observation that a variety of different patterns can be obtained with slight 
changes in the parameters points to other mechanisms controlling the growth. 
In particular, the spiral patterns are reminiscent of those observed in some liquid phase reaction-diffusion 
systems, such as the Belousov-Zhabotinsky reaction \cite{Markus1994, Perez1991, Moller1987}. 
In these systems, the instability is driven by an autocatalytic reaction. But spirals, and the zig-zag and oscillatory patterns visible 
in Fig. \ref{fig:scary_pattern}, are also characteristic of crack propagation in thin films \cite{Marthelot2014, Yuse1993, Fineberg1991}. 
Our patterns similarly arise spontaneously from a nonlinear reaction - diffusion type mechanism. 
However in our case, the growth mechanism 
involves the interplay of chemical etching and mechanical stress, which deforms the metal film catalyst. 
The Ge - Cr interface may be pre-stressed for several reasons, including incommensurate lattice constants of Ge and Cr, 
and different thermal expansion coefficients of Ge and the Cr film. 
Assuming the metal film is under compressive stress, if the Ge surface is etched away from under it, the metal film may buckle, 
leaving a void into which etching solution and products can diffuse. The buckling instability 
of the metal film may be the mechanism which selects the wavelength of the radial pattern, as we discuss below. 
Conversely, if the metal film is under tensile stress, it may crack, facilitating diffusion of the etch solution through the crack. \\ 
Etching of Ge proceeds through the formation of Germanium oxide ($Ge O_2$). In our system, a possible sequence 
of overall reactions is \cite{Kawase2013}: 
\begin{equation}
	Ge + 2 H_2 O \rightarrow Ge O_2 + 4 H^+ + 4 e^- 
	\label{eq:etch_reaction_ox}
\end{equation}
\noindent at the Ge - metal - water contact line, and 
\begin{equation}
	O_2 + 4 e^- + 4 H^+ \rightarrow 2 H_2 O
	\label{eq:etch_reaction_red}
\end{equation}
\noindent at the metal - water interface, for a net reaction $Ge + O_2 \rightarrow Ge O_2$ . Ge oxide is somewhat soluble 
in water, according to $Ge O_2 + 2 H_2 O \rightarrow Ge (O H)_4$ . Note that while this is 
a possible ``overall'' reaction sequence, the actual reaction steps may be more. 
Following the Ge oxidation reaction (\ref{eq:etch_reaction_ox}) the metal layer may play the role of an electron transporter, 
delivering electrons to the solution for the oxygen reduction reaction (\ref{eq:etch_reaction_red}). 
The Gold islands, which are necessary to generate the patterns, may provide a conductive path across the oxide layer 
which likely forms at the bare Cr - water interface, as well as providing a catalytic surface for the reaction (\ref{eq:etch_reaction_red}). 
Mechanisms of charge and mass transport are discussed in \cite{Geyer2012}. \\ 
Our patterns are associated with the buckling of the metal 
layer following its local detachment (``delamination'') from the Ge surface. 
Fig. \ref{fig:interference} is an interference 
microscopy picture of a radial pattern before the metal film lifts off the Ge surface. The interference fringes created 
by reflection from the semitransparent metal layer and the Ge surface show the buckling of the metal film; 
this radial pattern results in the etching pattern on the Ge surface shown in Fig. \ref{fig:Radial}. The tip splitting process 
of Fig. \ref{fig:Radial} is also visible in this interferogram. \\ 
Etching removes material from the Ge surface, creating a delamination area. Buckling of the metal 
inhibits further etching of the delaminated area (since in the void the metal catalyst is removed from the Ge surface), except 
at the Ge - metal - liquid contact line, where etching is catalyzed. Depending on the balance of forces on it, 
the contact line may move laterally, but also into the Ge surface \cite{Rykaczewski2011}, creating grooves. 
Buckling also facilitates diffusion of reactants to and from the contact line, and thus affects the reaction speed.  \\
The pattern of Fig. \ref{fig:Radial} exhibits a characteristic wavelength which is maintained by tip splitting 
as the pattern grows radially (Supp. Mat. Fig. S1). Assuming that the observed wavelength $\ell$ corresponds to 
the critical point of the Euler buckling instability for the metal layer, we can calculate the original strain in the metal. 
For a thin film of thickness $h$ under uniaxial compressive stress, the relation between 
critical strain $\epsilon$ and critical length for buckling $\ell$ is approximately (Supp. Mat.): 

\begin{equation}
	\epsilon = \frac{2 \pi^2}{3} \frac{h^2}{(1- \nu ^2) \, \ell ^2}
	\label{eq:critical_strain}
\end{equation}

\noindent where $\nu$ is the Poisson ratio. For Cr, $\nu \approx 0.2$ so $(1-\nu ^2) \approx 1$ ; 
with $h = 20 \, nm$ (metal film thickness) 
and $\ell = 20 \, \mu m$ (observed wavelength) we obtain $\epsilon \approx 6.6 \times 10^{-6}$ for the strain. 
This level of strain could easily result from the different thermal expansion of Ge and the metal film.  
The thermal expansion coefficient for Ge is $\alpha_{Ge} \approx 6 \times 10^{-6} \, ^\circ K^{-1}$ , 
while for (bulk) Cr it is in the range $\alpha_{Cr} \approx 4.5 - 5.5 \times 10^{-6} \, ^\circ K^{-1}$ \cite{ASM, White1986}. 
As an order of magnitude, the mismatch in thermal expansion coefficients 
in our system is therefore $\Delta \alpha \sim 10^{-6} \, ^\circ K^{-1}$. The strain measured from 
the buckling instability $\epsilon \approx 7 \times 10^{-6}$ would then correspond to a temperature difference 
$\Delta T = \epsilon / \Delta \alpha \sim 7 \, ^\circ C$. The sample of Fig. \ref{fig:interference} was incubated at $4 \, ^\circ C$, 
other samples at room temperature but without controlling the temperature. Thus the metal 
layer may be under compressive stress for some samples and under tensile stress for others. \\ 
In conclusion, we have presented a new pattern forming system which produces remarkable structures. While some 
of the patterns are reminiscent of reaction - diffusion systems, others recall patterns of crack propagation, 
or even thermal convection. Then there are similarities with forms in the living world. 
But this system is essentially solid state. Here, the mechanism for instability and pattern formation involves the interaction 
of a catalytic reaction with the mechanical deformation of the catalyst (the metal film). This mechanism has been exploited 
in some applications of MACE, by designing catalyst patches which fold into 3D shapes 
during the etching process \cite{Rykaczewski2011}, and by designing 2D catalyst shapes that induce 
chiral etching patterns \cite{Hildreth2012}. It is interesting to note that, on a different scale, 
mechano - chemical coupling is also the basis for the workings of enzymes, the molecular machines of life 
\cite{Molecular_Machines_Book}. Coincidentally, the overall reaction (\ref{eq:etch_reaction_ox}), (\ref{eq:etch_reaction_red}) 
parallels the overall reaction of aerobic respiration, with Germanium in the role of carbohydrates ! A theory for our patterns 
must describe the dynamics of the Ge - metal - liquid contact line, a challenging free boundary problem for future study. 
Similarly, identifying the experimental parameters which control growth and morphology of the patterns is a challenge for future 
experiments. Finally, the intersection with technologically important problems in materials science, such as
device fabrication, but also corrosion and crack propagation, surely merits continued attention. \\

\bibliography{Yilin_3_Archives}

\end{document}


\title{Supplementary Material for: \\
Metal Assisted Chemical Etching patterns at a Ge/Cr/Au interface \\ 
modulated by the Euler instability}
\author{Yilin Wong}
\author{Giovanni Zocchi}

\maketitle

\section{Materials and Methods}

\noindent {\bf Sample preparation.} The polished p-doped Ge wafer, $127 \, \mu m$ thick, $4 \, inch$ diameter, 
with [100] surface crystal orientation, is $O_2$ plasma cleaned for $10 \, min$. Immediately afterwards the metal layers are 
deposited on the Ge surface by electron beam evaporation:  $4 - 20 \, nm$ of Cr, depending on the sample, and $4 \, nm$ of Au. 
While the Cr layer is substantially uniform, the Au will typically form $\sim 50 \, nm$ size islands \cite{Yilin_2}. The wafer thus 
prepared is stored in air. For the experiments, the Ge-metal wafer is cut into $\sim \, 2 \, cm \times 4 \, cm$ rectangular pieces, 
which are rinsed with Acetone, Ethanol and D.I water sequentially to remove potential surface contamination, 
and blow dried with nitrogen. $500 \, \mu L$ of ``etching solution'' is then transferred on the surface of the chip, forming a drop 
covering part of it. The sample is now incubated at room temperature in open air, allowing the drop to evaporate. 
Within about 8 hours the sample is dry, with salt crystals from the etching solution covering the surface.
The crystallized salt is rinsed off with D.I water and the chip blow dried. A small number of spiral patterns or etch pits may form 
and become visible under the microscope at this step, but the bulk of the patterns are generated during the next incubation step.
In this second step, the chip is re-incubated with another $500\mu L$ of etching solution for $24 - 48$ hours, in a moist chamber 
to prevent evaporation of the drop. During this time the patterns shown in the figures form. After blow drying, we mount 
the chip on a microscope slide for easy handling and storage. \\ 

\noindent {\bf Etching solution.} We use a mild etching solution, consisting of $1 \, M$ phosphate buffer ($KH_2PO_4$) in DI water, 
$pH 4$, with the addition of $10 \, \mu M$ of a 20 bases single stranded DNA oligomer (Solution 1). The DNA is thiol modified 
at one end, for covalent attachment to the gold. It's presence is ``historical'': the initial patterns were obtained (accidentally) 
with this formulation, so we kept it in subsequent experiments. However, later experiments showed that the presence of DNA 
is not obligatory for forming the patterns. At the same time, we cannot exclude that DNA attached to the gold may participate 
in the reaction or the catalysis, moving the system in parameter space somewhat. For example, we know that adding a 
reducig agent to the etching solution has a dramatic effect on the patterns (Fig. 2). For DNA, its presence 
appears to speed up the etching process. \\ 
For some experiments, we used the same etching solution as above, without the DNA, and with the addition of 
$10 \, mM$ TCEP (a reducing agent): $1 \, M$ phosphate buffer ($KH_2PO_4$) in DI water, $pH 4$, with $10 \, mM$ TCEP 
(Solution 2).  \\

\noindent {\bf Visualization.} The patterns are observed with a metallurgical microscope (i.e. with white light illumination 
through the objective) at relatively low magnification ($5 \, \times$, $10 \, \times$, or $20 \, \times$ objective), and recorded 
with a digital camera attached to the microscope. Time lapse movies of the growing patterns are also recorded this way. \\ 
With the wet sample and the same setup we can observe corrugations of the metal layer (before it lifts off the Ge surface 
completely) by interference microscopy (Fig. 4): the interference fringes are created by reflection 
from the semi transparent metal layer and the high reflectivity Ge surface underneath. \\ 
SEM images of etch pitts and spiral patterns were obtained with a ZEISS Supra 40VP SEM 
with a secondary electron detector at 10kV. \\
All figures in the paper correspond to samples incubated at room temperature, with the exception of Fig. 4, where 
the sample was incubated at $4 ^\circ C$. \\

\noindent Finally, we note that formation of the patterns we describe in this report requires a combination of conditions and elements 
in a potentially high dimensional parameter space which we have by no means explored. Important parameters include 
the nature of the substrate (e.g. Ge vs Si), the specific surface (e.g. [100] vs [110]), the thickness of the metal layers, 
the chemical composition of the etching solution, the temperature, and possibly temperature gradients originating from 
the evaporation of the drop during day 1 of the incubation. The presence of both metal layers (Cr and Au) is  
essential. However a thick Au layer ($\ge 20 \, nm$) completely inhibits formation of the patterns (by being impermeable 
to the etching solution, presumably).  Although DNA is not obligatory, it does seem to expedite 
the pattern development process. \\

\section{Buckling of the metal layer} 

\noindent In this section, we derive eq. (3) of the paper. We consider the free energy of bending of a thin plate, in relation to 
the free energy for a uniaxial compression of the plate. The straight (not bent) plate lies on the xy plane; we consider 
a rectangular plate of thickness $h$, length $L$ along the x axis and $\ell$ along the y axis; $h << L , \ell$. We consider 
a uniaxial stress $\sigma$ along $y$, and the deformation of the plate to be uniform along $x$. Let $u(x, y)$ be the height 
of the (deformed) plate above $z = 0$ ; then with our conditions $\partial u / \partial x = 0$. For small deformations 
the elastic energy of bending is \cite{Landau_Lifshitz}: 

\begin{equation}
\begin{split}
	F = \frac{Y h^3}{24 \, (1 - \nu^2)} \iint dx dy [ ( \frac{\partial^2 u}{\partial x^2} + \frac{\partial^2 u}{\partial y^2} )^2 \\ 
	+ 2 (1 - \nu) \{ (\partial^2 u / \partial x \partial y)^2 - (\partial^2 u / \partial x^2) (\partial^2 u / \partial y^2) \} ]  
	\end{split}
	\label{eq:elastic_energy_1}
\end{equation}

\noindent where $Y$  is Young's modulus and $\nu$ the Poisson ratio; $h$ is the thickness of the plate. For the case 
of uniaxial compression along $y$, this formula becomes: 

\begin{equation}
	F_b = \frac{Y h^3}{24 \, (1 - \nu^2)} L \int_0^{\ell} dy \left ( \frac{\partial^2 u}{\partial y^2} \right )^2 
	\label{eq:elastic_energy_2}
\end{equation}

\noindent If the plate remains straight, and is compressed by an amount $\Delta y$ 
(i.e. $\epsilon = \Delta y / \ell$ is the strain), the elastic energy is: 

\begin{equation}
	F_s = \frac{1}{2} \, \frac{Y h L}{\ell} (\Delta y)^2 
	\label{eq:energy_straight}
\end{equation}

\noindent For the shape of the buckled plate we take the trial function: 

\begin{equation}
	u(y) = A \, \left [cos \left (\frac{2 \pi}{\ell} y \right ) + 1 \right ] \quad , \quad - \frac{\ell}{2} \le y \le \frac{\ell}{2}
	\label{eq:buckled_shape}
\end{equation}

\noindent which satisfies the right boundary conditions (zero curvature at the ends). This choice of shape amounts to 
using a third order polynomial (the lowest order which can satisfy the boundary conditions), but it turns out it gives the exact 
shape in the small bending approximation (see the discussion of the Euler instability in \cite{Molecular_Machines_Book}). 
The relation between the amplitude $A$ and $\Delta y$ is given by enforcing the contour length of the buckled plate: 

\begin{equation}
\begin{split}
	\ell = \int_{- \ell / 2 - \Delta y / 2}^{\ell / 2 + \Delta y / 2} dy \, \left [1 + \left (\frac{\partial u}{\partial y} \right )^2 \right]^{1/2} \\ 
	\approx \int_{- \ell / 2 - \Delta y / 2}^{\ell / 2 + \Delta y / 2} dy \, \left [1 + \frac{1}{2} \left (\frac{\partial u}{\partial y} \right )^2 \right ] 
	\end{split} 
	\label{eq:contour_length}
\end{equation}

\noindent Using the shape (\ref{eq:buckled_shape}), the condition (\ref{eq:contour_length}) gives: 

\begin{equation}
	|\Delta y| = A^2 \, \frac{\pi ^2}{\ell}
	\label{eq:buckled_shape}
\end{equation}

\noindent to lowest order in the small quantity $(\Delta y / \ell)$. Finally, the free energy of the bent plate is: 

\begin{equation}
	F_b = Y \, \frac{h^3 L}{24 \, (1 - \nu^2)} \frac{2 (2 \pi)^2}{\ell ^2} |\Delta y|
	\label{eq:energy_bent}
\end{equation}

\noindent while the free energy for the straight (compressed) plate is given by (\ref{eq:energy_straight}). 
Equating the two expressions we obtain the critical compression for buckling: 

\begin{equation}
	(\Delta y)_c = \frac{2}{3} \frac{h^2}{1- \nu ^2} \frac{\pi ^2}{\ell}
	\label{eq:critical_compression}
\end{equation}

\noindent and the corresponding critical strain: 

\begin{equation}
	\epsilon_c = \frac{(\Delta y)_c}{\ell} =  \frac{2 \pi ^2}{3} \frac{h^2}{(1- \nu ^2) \, \ell^2} 
	\label{eq:critical_strain}
\end{equation}

\section{Wavelength of the radial pattern}

\noindent Radial patterns as in Fig. 2 are obtained with the addition of $10 \, mM$ TCEP (Tris(2-carboxyethyl)phosphine hydrochloride: 
a reducing agent) to the etching solution. Presumably its effect is to slow down the Ge oxidation reaction. On the other hand, 
addition of an oxidizing agent ($H_2 O_2$) does not seem to have a dramatic effect under our conditions. 

\begin{figure}[H]
	\centering
	\includegraphics[width=\linewidth]{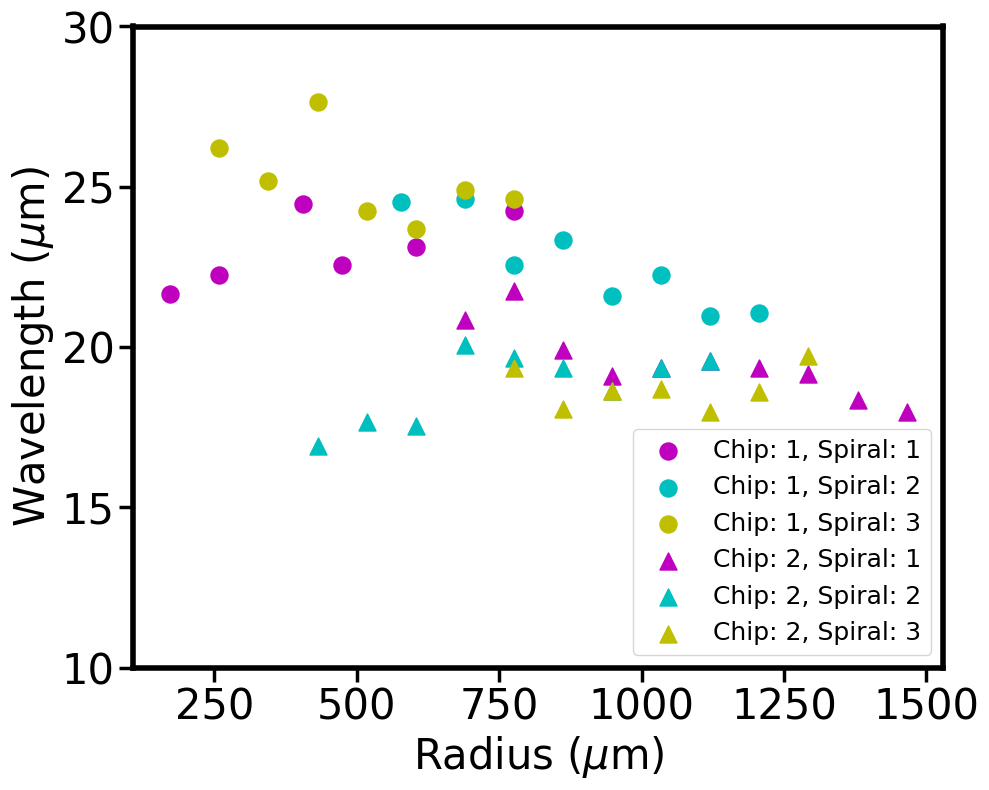}
	\caption{Wavelength $\ell$ (distance between grooves) of the radial pattern of grooves 
	measured at different radii $r$ (distance from the center) for 6 different radial patterns as in Fig. 2. 
	Different colors refer to different patterns on the same sample; circles and squares refer to two different samples. 
	The two samples were grown under nominally the same conditions, with $20 \, nm$ Cr 
	and $4 \, nm$ Au and TCEP (Solution 2). 
	The wavelength remains approximately constant (independent of $r$) for a given pattern, and is also essentially the same 
	for patterns on the same chip ($\ell \approx 24 \, \mu m$ and $\ell \approx 19 \, \mu m$ for the two samples shown). 
	As the grooves grow radially, the distance $\ell$ is maintained constant through a tip splitting instability of the grooves. }
	\label{fig:lambda_vs_r}
\end{figure}

\begin{figure}[H]
	\centering
	\includegraphics[width=3.2 in]{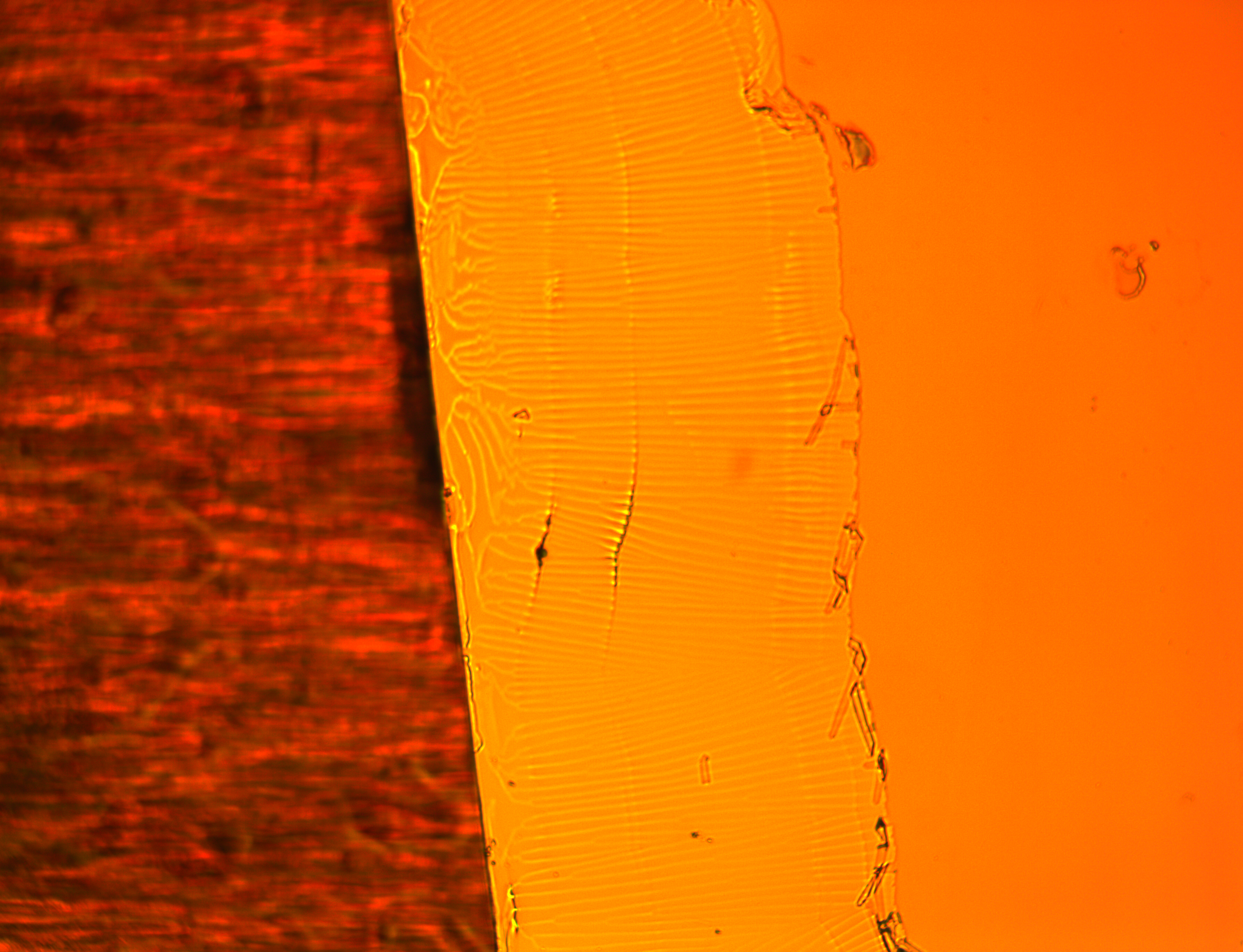}
	\caption{Parallel grooves growing perpendicular to a straight etch front which started from the edge of the sample. 
	The deeper orange area on the right has the metal film still attached to the Ge surface, while in the lighter orange 
	area the film has lifted off during the etching process. Shards of metal film are visible at the etch front.   }
	\label{fig:straight_etch_wrinkles}
\end{figure}

\noindent One notable feature of the radial pattern is the tip splitting process which maintains an approximately constant 
distance $\ell$ between grooves as the pattern grows radially. The tip splitting ``fork'' has a characteristic shape 
(Fig. 2). In Fig. S\ref{fig:lambda_vs_r} we display measurements of the wavelength $\ell$ at varying distance $r$ from 
the center of the patterns, showing that indeed $\ell$ is approximately independent of $r$. 
We associate the wavelength of these radial patterns with the buckling instability of the metal layer. \\ 
\noindent Fig. S\ref{fig:straight_etch_wrinkles} shows the etching pattern resulting from the buckling instability 
of the metal layer, in the case of a straight etching front. Here etching started from the edge of the sample,  
visible in the picture, instead of starting from an etch pit. The result is a pattern of straight parallel lines 
(and thus no tip splitting), instead of a pattern of radial lines. The wavelength of the pattern 
in Fig. S\ref{fig:straight_etch_wrinkles} is $\ell \approx 20 \, \mu m$. With this value, and the metal film parameters 
($h = 20 \, nm $ , $(1 - \nu ^2) \approx 1$ for Cr), we obtain from (\ref{eq:critical_strain}) a value for the strain 
in the pre-stressed metal film: $\epsilon \approx 6.6 \times 10^{-6}$ . 
We remark in the main text that this level of strain could result from the different thermal expansion of Ge and the metal film.  
The thermal expansion coefficient for Ge is $\alpha_{Ge} \approx 6 \times 10^{-6} \, ^\circ K^{-1}$. The precise 
value for Cr, especially for a thin film, is more delicate. Cr has an antiferromagnetic transition at a Neel temperature 
$T_N \approx 311 \, ^\circ K $ \cite{White1986}; associated with this phase transition, the thermal expansion coefficient 
has a non monotonic behavior close to room temperature. For bulk Cr around room temeperature, the thermal expansion 
coefficient is in the range $\alpha_{Cr} \approx 4.5 - 5.5 \times 10^{-6} \, ^\circ K^{-1}$ 
\cite{ASM, White1986, Koumelis1973}. 

\section{Spirals and pitts}

\begin{figure}[H]
	\centering
	\includegraphics[width=\linewidth]{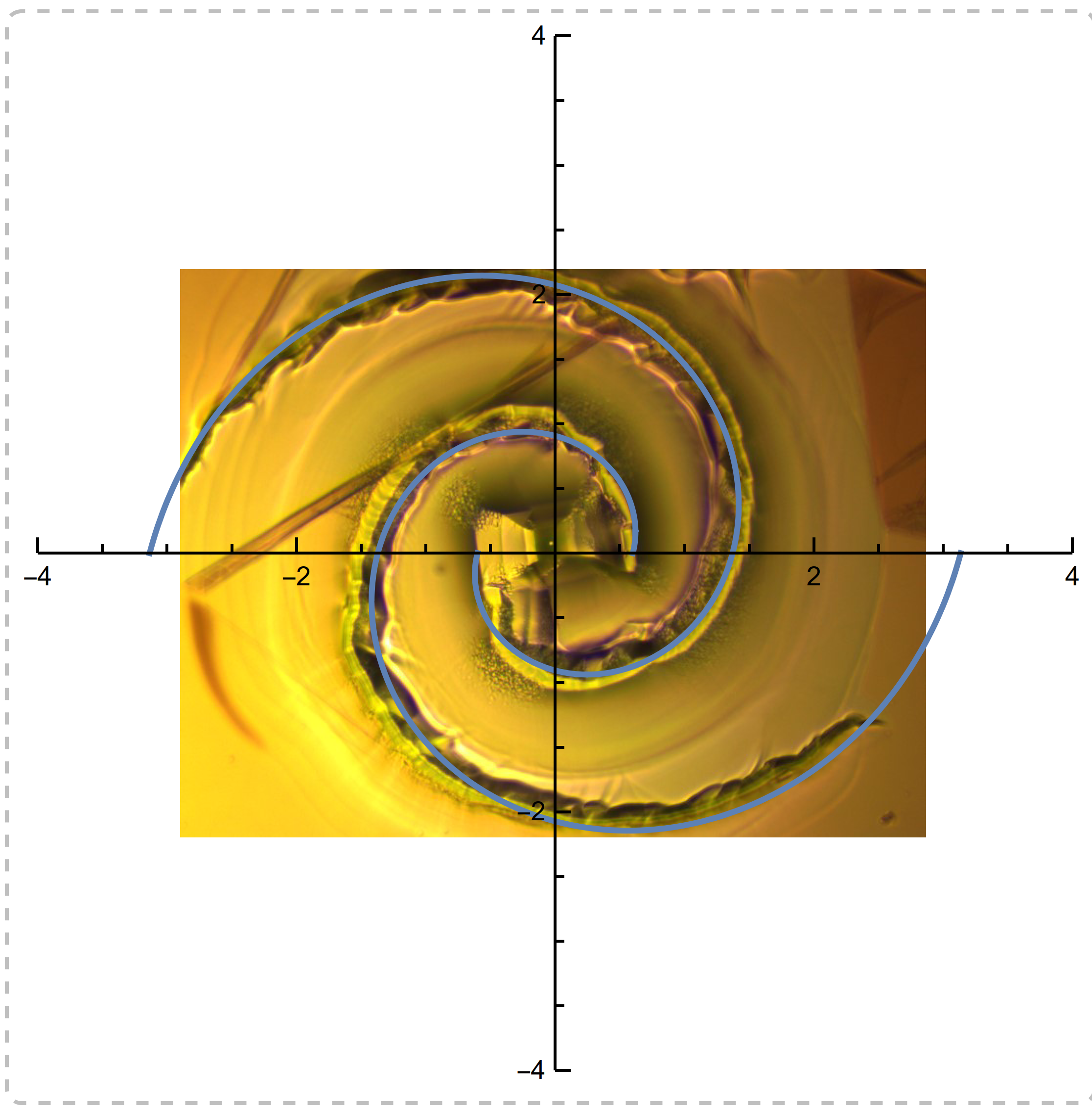}
	\caption{The spiral pattern of Fig. 1b with superimposed fit to a logarithmic spiral (blue line). The two arms are 
	 $r_1 (\theta) = r_0 e^{\theta / {(\omega \tau)}}$ and $r_2 (\theta) = r_0 e^{(\theta - \pi) / {(\omega \tau)}}$ 
	 with $(\omega \tau) = 3.8 $. }
	\label{fig:Log_spiral_fit}
\end{figure}

In parametric form, the logarithmic spiral is $r(t) = r_0 e^{t / \tau}$ ; $\theta (t) = \omega t$, or 
$r(\theta) = r_0 e^{\theta / {(\omega \tau)}}$. Fig. S\ref{fig:Log_spiral_fit} shows this form superimposed on 
an experimental pattern. 

\begin{figure}[H]
	\centering
	\includegraphics[width=\linewidth]{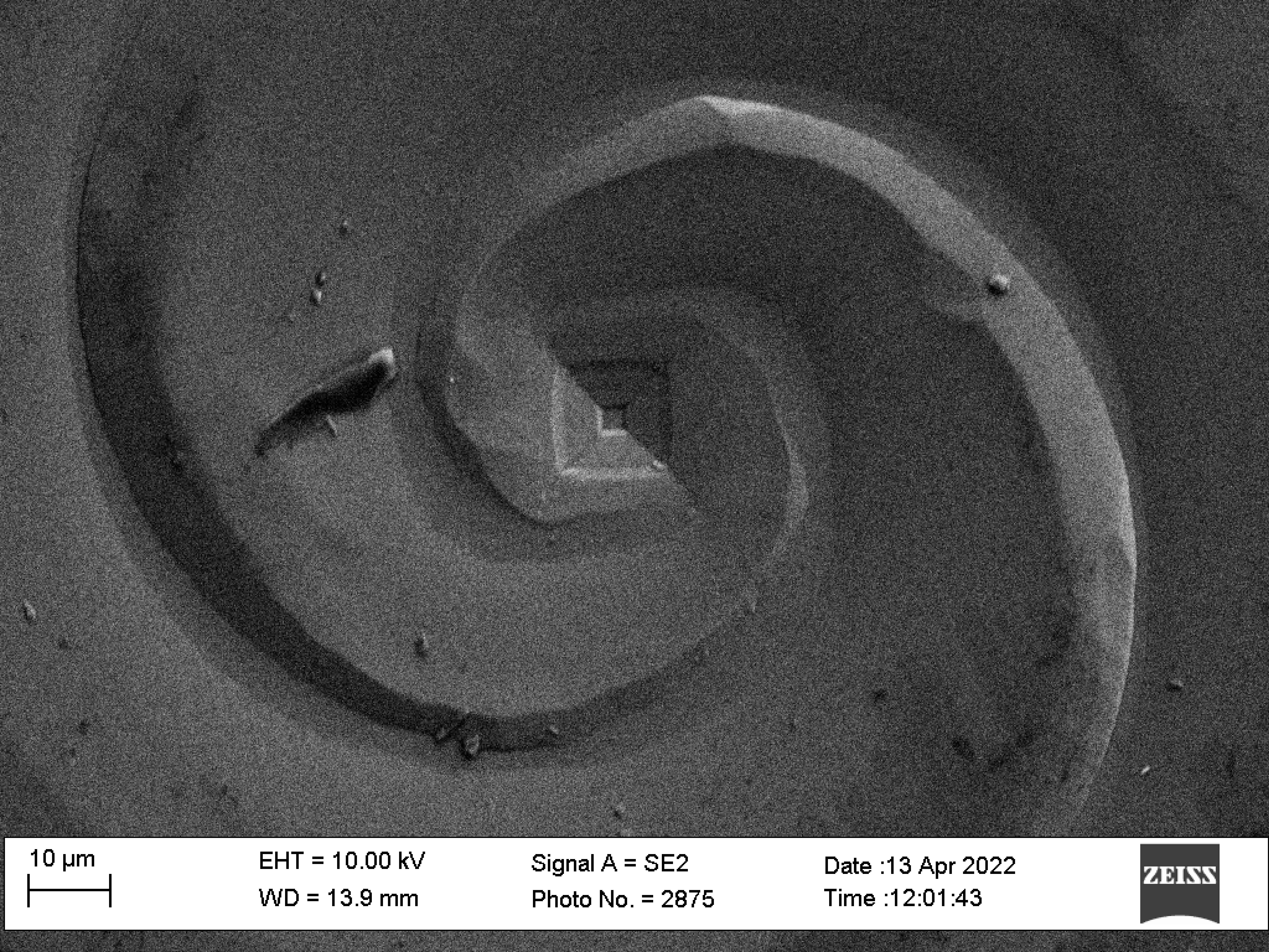}
	\caption{SEM picture of a logarithmic spiral. The two arms of the spiral grow out of the square etch pit 
	at the center, which has the shape of an inverted truncated pyramid. }
	\label{fig:SEM_spiral}
\end{figure} 

\begin{figure}[H]
	\centering
	\includegraphics[width=\linewidth]{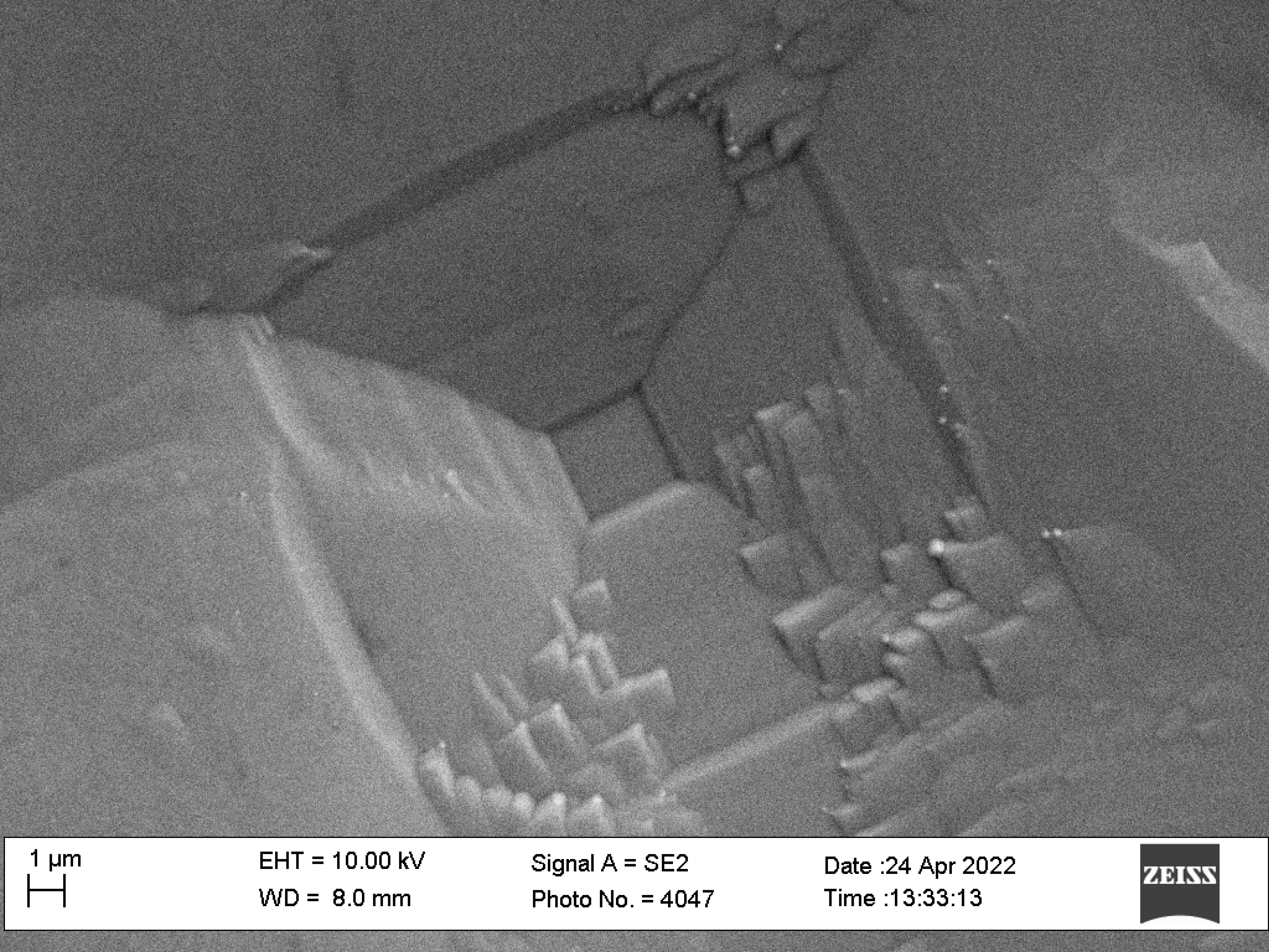}
	\caption{SEM picture of an etch pit showing steps and facets at the $\mu m$ scale. }
	\label{fig:SEM_etch_pit}
\end{figure}

\section{Growth dynamics}

\begin{figure}[H]
	\centering
	\includegraphics[width=\linewidth]{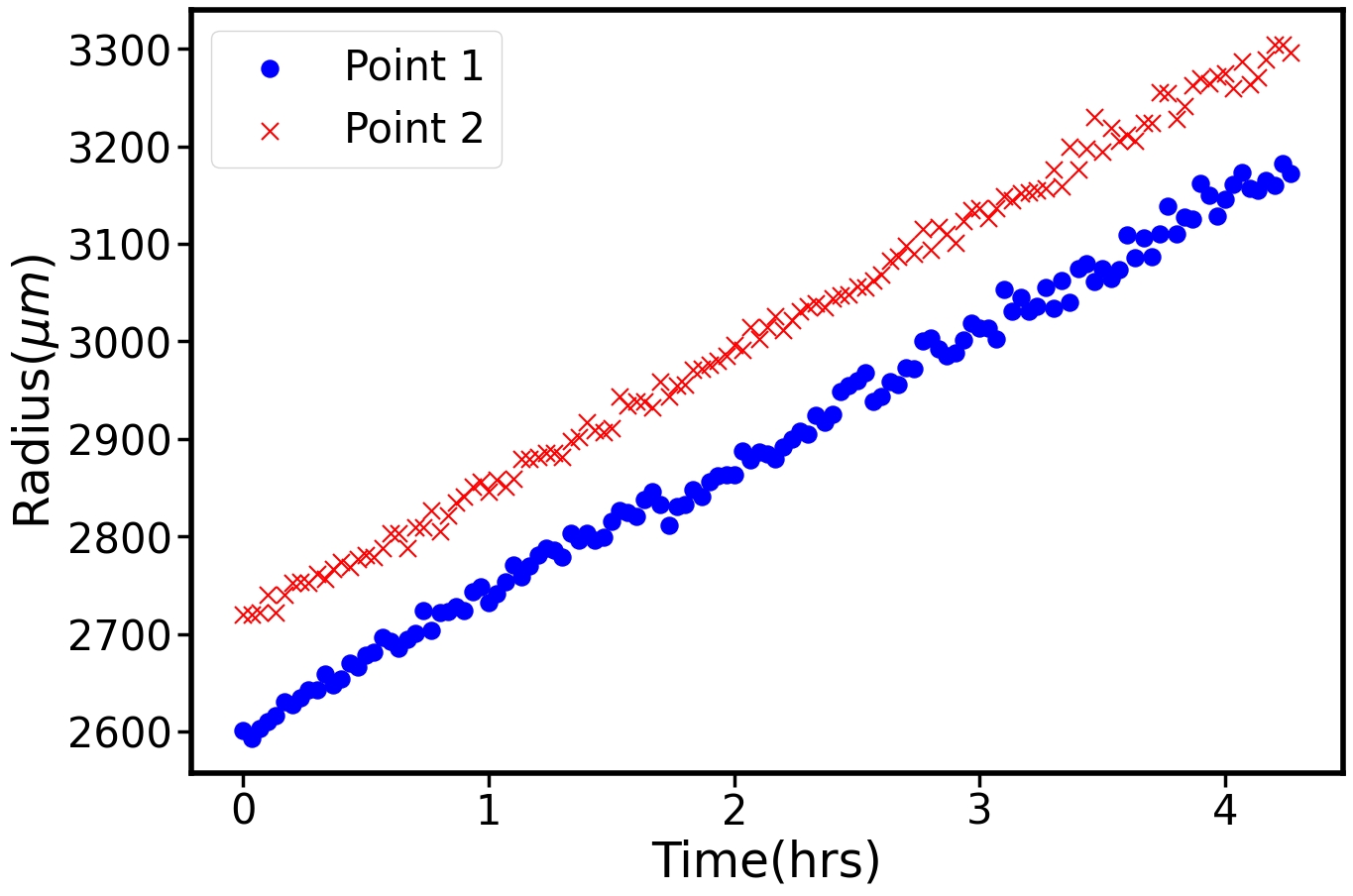}
	\caption{Radial growth ($r$ vs $t$) of a radial pattern extracted from video microscopy. The two different colors represent 
	two different points at the edge of the expanding pattern, diametrically opposed. The conditions were 
	surface deposition of $20 \, nm$ Cr and $4 \, nm$ Au and incubation with TCEP (Solution 2). 
	The growing disk is not perfectly circular; its ``center'' was determined from the final pattern. 
	For the two points chosen for the plot, the distance to this center differs by $100 \, \mu m$; nevertheless 
	the overall radial growth rate stays constant at $dr / dt \approx 2.5 \, \mu m/min$ . }
	\label{fig:Radial_Radius}
\end{figure}

\begin{figure}[H]
	\centering
	\includegraphics[width=\linewidth]{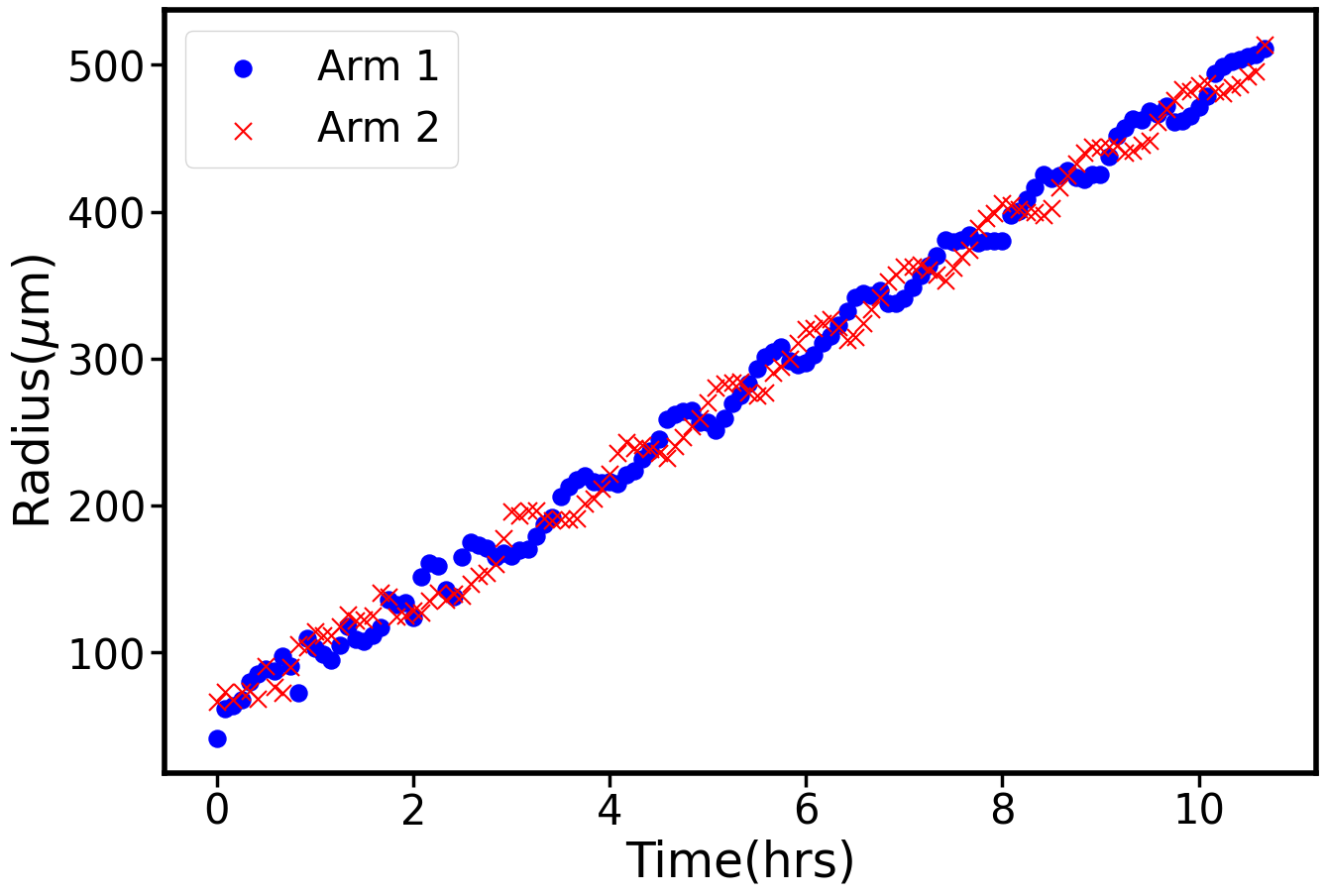}
	\caption{Radial growth ($r$ vs $t$) of a two arms Archimedean spiral extracted from video microscopy. 
	The two different colors represent the two arms of the spiral. Specifically, the blue dots 
	are the radial position of the tip of one arm of the spiral, in the course of time. 
	The red crosses are the radial position of the tip of the other arm. 
	Conditions were $20 \, nm$ Cr and $4 \, nm$ Au and DNA (Solution 1). 
	The overall slope of the plot is $dr / dt \approx 0.75 \, \mu m/min$. The periodic modulation 
	visible in the plot is caused by an external field gradient, namely nonuniform illumination under the microscope 
	(see text).}
	\label{fig:Spiral_Radius}
\end{figure}

\begin{figure}[H]
	\centering
	\includegraphics[width=\linewidth]{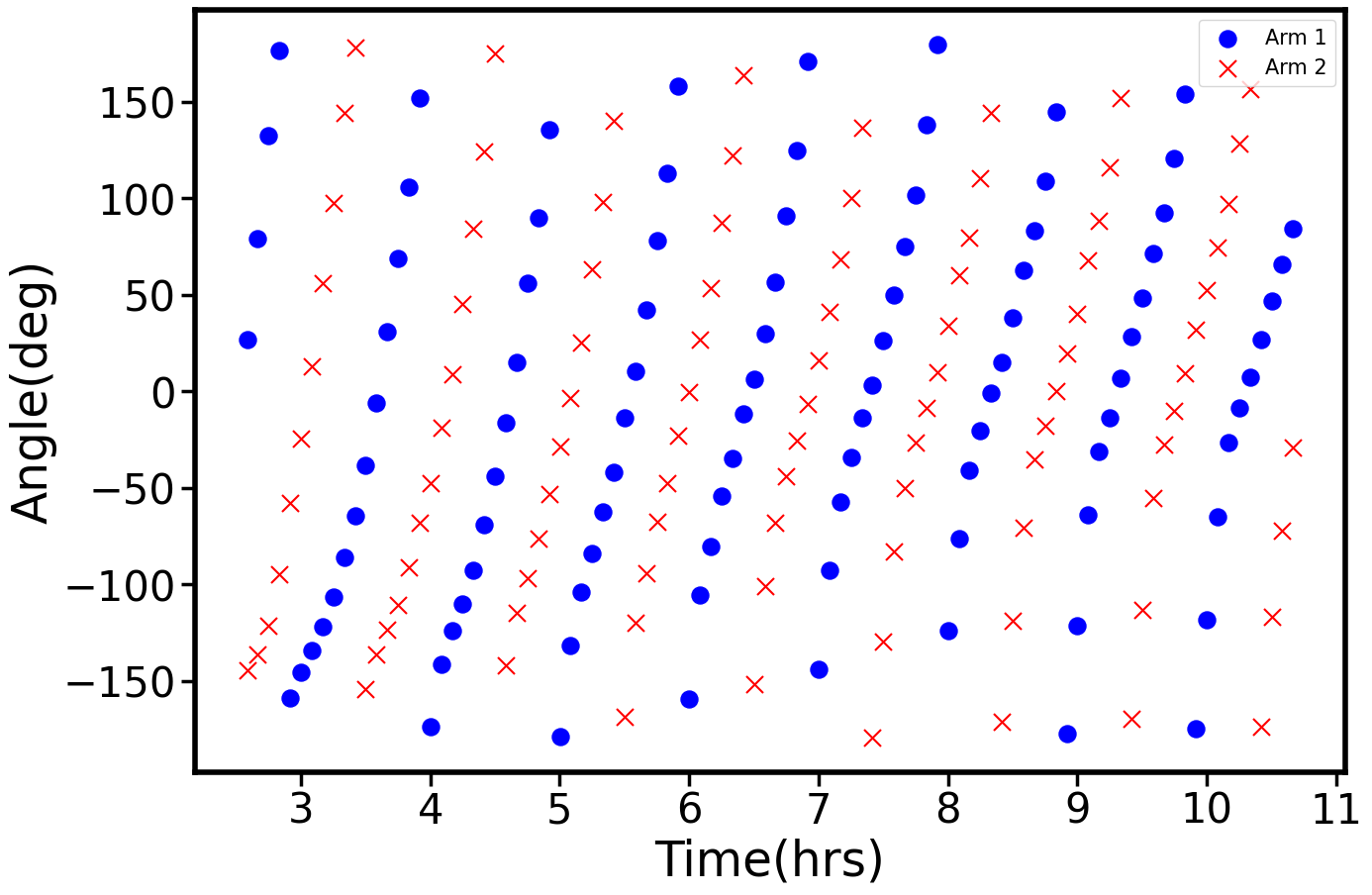}
	\caption{For the same spiral as in Fig. S\ref{fig:Spiral_Radius}, this plot displays the angle $\theta$ 
	corresponding to the position of the tip of the spiral arm, in polar coordinates. The angle is plotted in degrees 
	in the interval $- 180 \, ^{\circ} < \theta \le 180 \, ^{\circ}$, modulo $360 \, ^{\circ}$ . The blue dots correspond 
	to one spiral arm and the red crosses to the other. On average the angular velocity $d \theta / dt$ is constant in 
	the course of time, but there is a periodic modulation visible in the slope of the lines in the plot. Namely, within the sector 
	$- 50 ^{\circ} < \theta < 50 ^{\circ} $ , $d \theta / dt$ slows down compared to within the sector 
	$ 100 ^{\circ} < \theta < 200 ^{\circ} = - 160 ^{\circ}$, where $d \theta / dt$ speeds up. }
	\label{fig:Spiral_Angle}
\end{figure}

\begin{figure}[H]
	\centering
	\includegraphics[width=\linewidth]{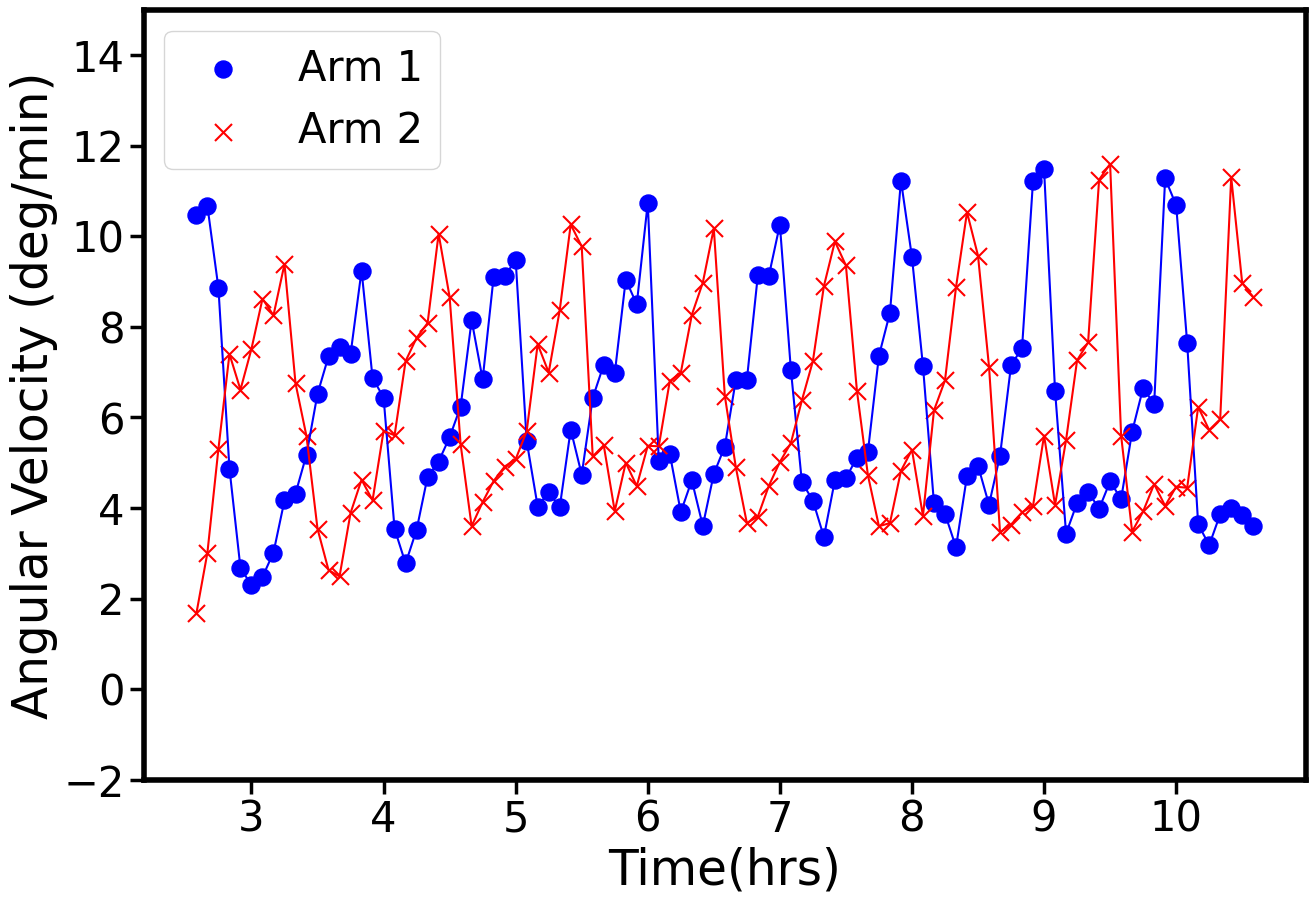}
	\caption{Angular velocity $d \theta / dt$ of the tip of the spiral arms. This plot is constructed 
	with the same data as in Fig. S\ref{fig:Spiral_Angle}. It can be seen that the angular velocity of the spiral 
	averages around $< d \theta / dt > \, \approx 6 \, deg / min$ , with a significant modulation discussed in the text. 
	The same modulation can be seen in Fig. S\ref{fig:Spiral_Angle} as a variation in the slope of the lines in each cycle. 
	It corresponds to regions of different brightness in the field of view of the microscope.}
	\label{fig:Spiral_AngularVelocity}
\end{figure}

\noindent Through time-lapse video microscopy we were able to extract data on the growth dynamics of the Archimedean spirals and Radial patterns. Both the Archimedean spiral and the radial pattern have constant radial growth rate. 
For conditions of $20 \, nm$ Cr and $4 \, nm$ Au, the growth rate of the radius for the Archimedean spiral is 
$dr / dt \approx 0.75 \, \mu m / min$ (Fig. S\ref{fig:Spiral_Radius}) and the growth rate of the radius of the radial pattern 
is $dr / dt \approx 2.5 \, \mu m / min$ (Fig. S\ref{fig:Radial_Radius}). The periodic modulation visible in 
Fig. S\ref{fig:Spiral_Radius} is addressed below. \\ 
The Archimedean spiral grows at a relatively constant angular velocity. 
In Fig. S\ref{fig:Spiral_Angle} we plot the angle corresponding to the position, in polar coordinates, 
of the tip of the growing arms, for a two - arms Archimedean spiral. The angle $\theta$ is plotted modulo $360 \, ^{\circ}$ 
in the interval $- 180 \, ^{\circ} < \theta \le 180 \, ^{\circ}$ . A periodic modulation of the angular velocity is 
noticeable, also displayed in Fig. S\ref{fig:Spiral_AngularVelocity}, where we plot the derivative $d \theta / dt$ 
of the data in Fig. S\ref{fig:Spiral_Angle}. The modulation in angular velocity corresponds to an uneven illumination 
field under the microscope, which happened to be brighter on one side of the field of view. The angular velocity 
was correspondingly faster on that side, and slower on the dimmer side. This is also the origin of the modulation visible 
in Fig. S\ref{fig:Spiral_Radius}. We are not yet sure about the physical origin 
of this effect: it could be related to heating of the sample, or to a photochemical mechanism. In any case this is 
an interesting phenomenon from the point of view of identifying control parameters for the growth, and merits further 
study.  \\

\bibliography{Yilin_3_Supp_Mat}